\documentclass{llncs}
\usepackage[T1]{fontenc}
\usepackage{graphicx}
\usepackage{wrapfig}
\usepackage{lipsum}
\usepackage{caption}
\usepackage{microtype} 
\usepackage{rotating}
\usepackage{multirow}
\usepackage{makecell}
\usepackage{adjustbox}
\usepackage{float}
\usepackage{cite}
\usepackage{textcomp}
\usepackage{hyperref}
\usepackage{amsmath}
\hypersetup{
    colorlinks=false, 
    linkcolor=black, 
    citecolor=black, 
    urlcolor=black, 
    pdfborder={0 0 0} 
}

\title{Green Recommender Systems: Optimizing Dataset Size for Energy-Efficient Algorithm Performance}
\author{Ardalan Arabzadeh\inst{1} \and
Tobias Vente\inst{1} \and
Joeran Beel\inst{1,2}}
\authorrunning{Arabzadeh et al.}
\institute{University of Siegen, Department of Electrical Engineering and Computer Science, Germany \and
Recommender-Systems.com, Siegen, Germany\\
\email{joeran.beel@uni-siegen.de}\\
\url{https://www.recommender-systems.com} } 

\begin{document}

\maketitle

\begin{abstract}
As recommender systems become increasingly prevalent, the environmental impact and energy efficiency of training these large-scale models have come under scrutiny. This paper investigates the potential for energy-efficient algorithm performance by optimizing dataset sizes through downsampling techniques. We conducted experiments on the MovieLens 100K, 1M, 10M and Amazon Toys and Games datasets, analyzing the performance of various recommender algorithms under different portions of dataset size. Our results indicate that while more training data generally leads to higher performance in algorithms, certain algorithms, such as FunkSVD and BiasedMF, particularly in cases involving more unbalanced and sparse dataset like Amazon Toys and Games, maintain high-quality recommendations with up to 50\% reduction in training data, achieving nDCG@10 scores within $\sim$13\% of their full dataset performance. These findings suggest that strategic dataset reduction can decrease computational and environmental costs without substantially compromising recommendation quality. This study advances sustainable and green recommender systems by providing actionable insights for reducing energy consumption while maintaining effectiveness.\\

\newline


\end{abstract}

\section{Introduction}
Advancements in recommender systems have enhanced user experience. However, these advancements came at a substantial computational and energy cost \cite{vente2024clicks,spillo2023sustainability}. Large datasets do not only increase operational expenses but also result in higher energy consumption and carbon emissions, contributing to a more significant environmental impact \cite{spillo2023sustainability,vente2024clicks,al-jarrah2015efficient,chen2020deep,Spillo2024}. In extreme cases, energy consumption between datasets differs by factor 1,444, such as between LastFM vs. Yelp with the DGCF algorithm \cite{vente2024clicks}.  

Given the environmental and computational challenges associated with large datasets, it's important to question whether using the entire datasets is always necessary. For instance, datasets like MovieLens 10M are frequently employed in training recommender systems. But is it necessary to use 10 million instances, especially with simple baseline algorithms? Or would downsampling the dataset to e.g. 10\% suffices, which, in turn, might save 90\% of energy? 

In our work, we investigate whether downsampling the dataset can lead to an acceptable trade-off between energy efficiency and the performance of recommender algorithms. We see our work in the context of "Green Recommender Systems" as defined by Beel et al. as follows \cite{Beel2024e}.

\begin{quote}
\textit{``Green Recommender Systems'' are recommender systems designed to minimize their environmental impact throughout their life cycle - from research and design to implementation and operation. Green Recommender Systems typically aim to match the performance of traditional systems but may also accept trade-offs in accuracy or other metrics to prioritize sustainability. Minimizing environmental impact typically but not necessarily means minimizing energy consumption and CO\textsubscript{2} emissions.} \cite{Beel2024e}
\end{quote}

\noindent For our current work, we hypothesize that by downsampling recommender system datasets, we can save time, energy, and CO2 emissions, while obtaining nearly the same performance as with the full datasets.

\section{Related Work}
The field of green recommender systems has only started evolving recently \cite{spillo2023sustainability, vente2024clicks}. We also recently proposed "e-fold cross-validation", an energy-efficient alternative to k-fold cross-validation \cite{Mahlich2024,Baumgart2024}. Wegmeth et al. introduced EMERS, a tool to measure the electricity consumption of recommender system experiments \cite{Wegmeth2024a}. Also, judging based on the "accepted papers" list of the RecSoGood workshop, more related work is to be published very soon \cite{Plaza2024,Spillo2024a}.

While the "green" concept in recommender systems is new, other disciplines, like Automated Machine Learning, explore options to save energy for a longer time \cite{Santos2024,ALZOUBI2024143090,castellanos2024strategies,hennig2024leveragingautomlsustainabledeep,tornede2023towards,Castellanosgreenai,greenai2020}. 

In the domain of recommender systems, several studies have explored the impact of dataset size on the efficiency of algorithmic performance, which aligns the key focus of this study. Notably, Bentzer and Thulin explore the trade-off between accuracy and computational efficiency in collaborative filtering algorithms under limited data conditions \cite{bentzer2023}. They found that IBCF algorithm performs better in terms of accuracy with smaller datasets compared to SVD algorithm, while SVD outperforms IBCF in terms of speed and scalability with larger datasets. Their study highlights the performance differences between these two algorithms but does not address how other algorithms perform under similar constraints. This gap is relevant to our research, which seeks to evaluate a wider range of algorithms for optimizing both energy efficiency and performance.

Additionally, Jain and Jindal's review emphasizes that strategic sampling and filtering can enhance recommendation efficiency by improving computational speed and accuracy\cite{jain2023sampling}. However, their review lacks experimental validation of how these techniques impact algorithm performance with varying dataset sizes. Our study addresses this gap by empirically evaluating these effects on recommender systems. Judging based on the paper's title and abstract, Spillo et al. appear to have conducted research similar to ours \cite{Spillo2024}. However, at the time of conducting our research and writing our manuscript, the work of Spillo et al. was not yet publicly available but only announced on the ACM Recommender Systems conference website as an accepted paper.

It is worth mentioning that most papers and experiments focusing on downsampling and data efficiency are predominantly conducted in domains like (automated) machine learning, AI and computer vision \cite{Santos2024, ALZOUBI2024143090,castellanos2024strategies,hennig2024leveragingautomlsustainabledeep,tornede2023towards,Castellanosgreenai}, with more extensive research compared to the recommender systems domain. Moreover, studies within this broader field also corroborate the potential benefits of downsampling. Research by Zogaj et al. demonstrates that reducing dataset sizes can enhance both computational efficiency and predictive accuracy in genetic programming-based AutoML systems\cite{zogaj2021doing}. Their experiments show that downsampling large datasets can even in some cases result in better performance than using the full dataset, with shorter search times.

These studies underscore and evaluate the potential benefits of downsampling and its impact on model performance, but are not directly applicable to traditional recommender system algorithms, where such effects remain underexplored. 

\section{Methodology}
\subsection{Datasets \& Preprocessing}
We used four datasets for our experiment: \textbf{MovieLens 100K}, \textbf{MovieLens 1M}, \textbf{MovieLens 10M}, and \textbf{Amazon Toys and Games}.
The MovieLens datasets feature relatively balanced ratings across a scale from 1 to 5. In contrast, the Amazon Toys and Games dataset exhibits a skewed distribution, with $\sim$90\% of ratings concentrated in the 4 and 5 ranges.
The following preprocessing steps were applied to the datasets:
removal of duplicate rows, averaging duplicate ratings, and applying 10-core pruning to retain users and items with at least 10 interactions.

The dataset details before and after preprocessing are in Table~\ref{tab:dataset_info}.

\begin{table}[htbp]
\centering
\caption{Basic information of datasets before and after preprocessing}
\label{tab:dataset_info}
\resizebox{\textwidth}{!}{%
\begin{tabular}{|l|c|c|c|c|c|c|c|c|c|c|}
\hline
& \multicolumn{5}{c|}{\textbf{Before Preprocessing}} & \multicolumn{5}{c|}{\textbf{After Preprocessing}} \\ \hline
\textbf{Dataset} & \textbf{\#Users} & \textbf{\#Items} & \textbf{\#Interactions} & \makecell{\textbf{Avg.}\\\textbf{\#Int. per}\\\textbf{user}} & \makecell{\textbf{Avg.}\\\textbf{\#Int. per}\\\textbf{item}} & \textbf{\#Users} & \textbf{\#Items} & \textbf{\#Interactions} & \makecell{\textbf{Avg.}\\\textbf{\#Int. per}\\\textbf{user}} & \makecell{\textbf{Avg.}\\\textbf{\#Int. per}\\\textbf{item}} \\ \hline
MovieLens 100K & 943 & 1,682 & 100,000 & 106 & 59 & 943 & 1,152 & 97,953 & 103 & 85 \\ 
MovieLens 1M & 6,040 & 3,706 & 1,000,209 & 165 & 269 & 6,040 & 3,260 & 998,539 & 165 & 306 \\ 
MovieLens 10M & 69,878 & 10,677 & 10,000,054 & 143 & 936 & 69,878 & 9708 & 9,995,471 & 143 & 1029 \\ 
Amazon Toys and Games & 208,180 & 78,772 & 1,828,971 & 8 & 23 & 11,609 & 8,443 & 202,721 & 17 & 24 \\ \hline
\end{tabular}%
}
\end{table}

\subsection{Data Splitting and Downsampling}
\label{subsec:splitting_downsampling}
We applied a User-Based Split\cite{meng2020exploring}, with 10\% of each user's interactions randomly selected for the test set, 10\% for validation, and 80\% for training. The validation set was used for hyperparameter tuning, maintaining a comparable size between the validation and test sets to account for the impact of the training-to-validation/test ratio on results, as highlighted in prior research\cite{canamares2020offline}. The training set was downsampled to various proportions (10\%, 20\%, 30\%, up to 100\%) by randomly selecting different portions of each user's interactions. This approach ensures consistency in user representation across all sets while varying the number of interactions in the training set.

\subsection{Algorithms and Evaluation}
We trained the following algorithms on the downsampled training sets using the LensKit \cite{ekstrand2020lenskit} and RecPack \cite{recpack2022} libraries:

\begin{table}[H]
\centering
\caption{Information of algorithms used in our experiment}\label{tab:example}

\resizebox{\textwidth}{!}{ 
\begin{tabular}{|l|c|c|c|c|c|c|c|c|c|c|c|}
\hline
\textbf{Algorithms} & \rotatebox{45}{Bias} & \rotatebox{45}{Popular} & \rotatebox{45}{Random} & \rotatebox{45}{UserKNN} & \rotatebox{45}{ItemKNN} & \rotatebox{45}{BiasedMF} & \rotatebox{45}{FunkSVD} & \rotatebox{45}{Popularity} & \rotatebox{45}{ItemKNN} & \rotatebox{45}{SVD} & \rotatebox{45}{NMF} \\ \hline
\textbf{Library}   & LensKit & LensKit & LensKit & LensKit & LensKit & LensKit & LensKit & RecPack & RecPack & RecPack & RecPack \\ \hline
\end{tabular}
}
\end{table}

Performance was evaluated using the nDCG@10 metric, ensuring that all libraries adhered to an identical standard calculation logic to facilitate a fair comparison of results across algorithms\cite{schmidt2024evaluating}.

\section{Results \& Conclusion}
\begin{figure}[htbp] 
    \centering
    \begin{minipage}[b]{0.48\linewidth}
        \centering
        \includegraphics[width=\linewidth]{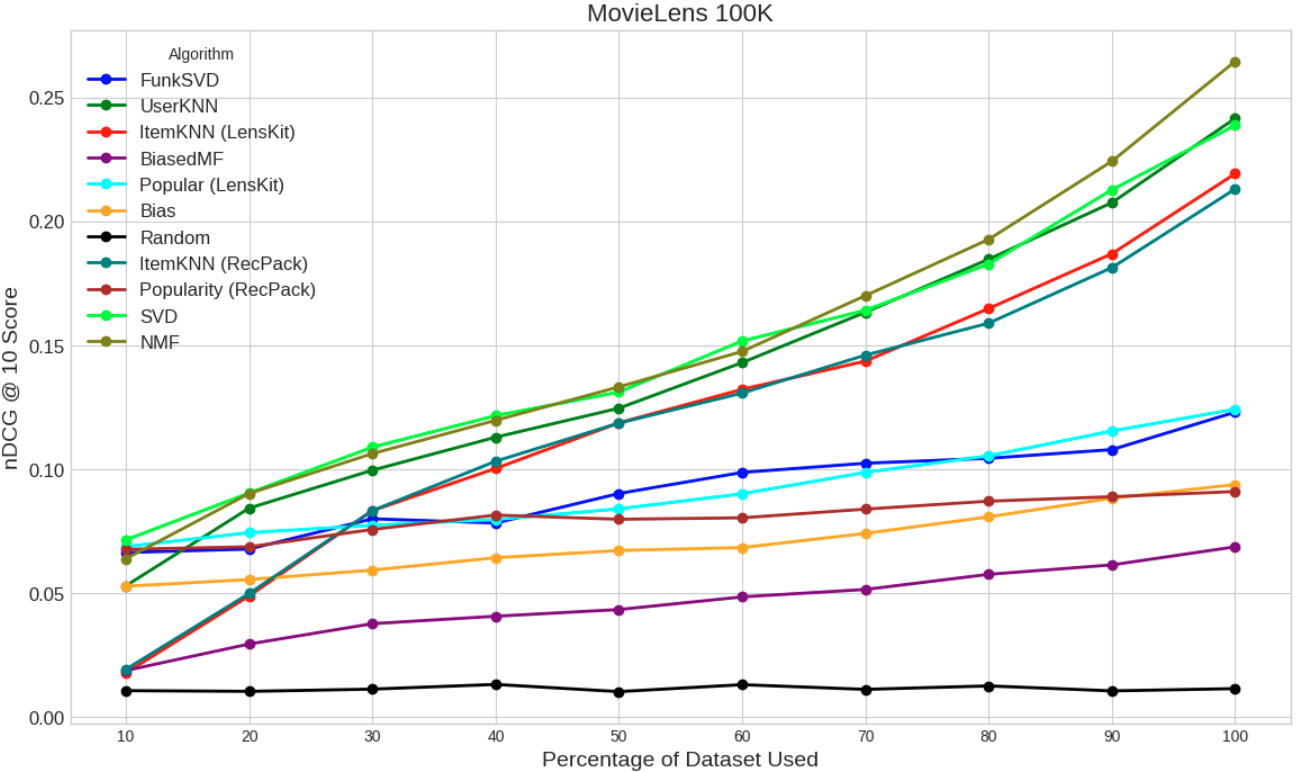}
    \end{minipage}%
    \hfill
    \begin{minipage}[b]{0.48\linewidth}
        \centering
        \includegraphics[width=\linewidth]{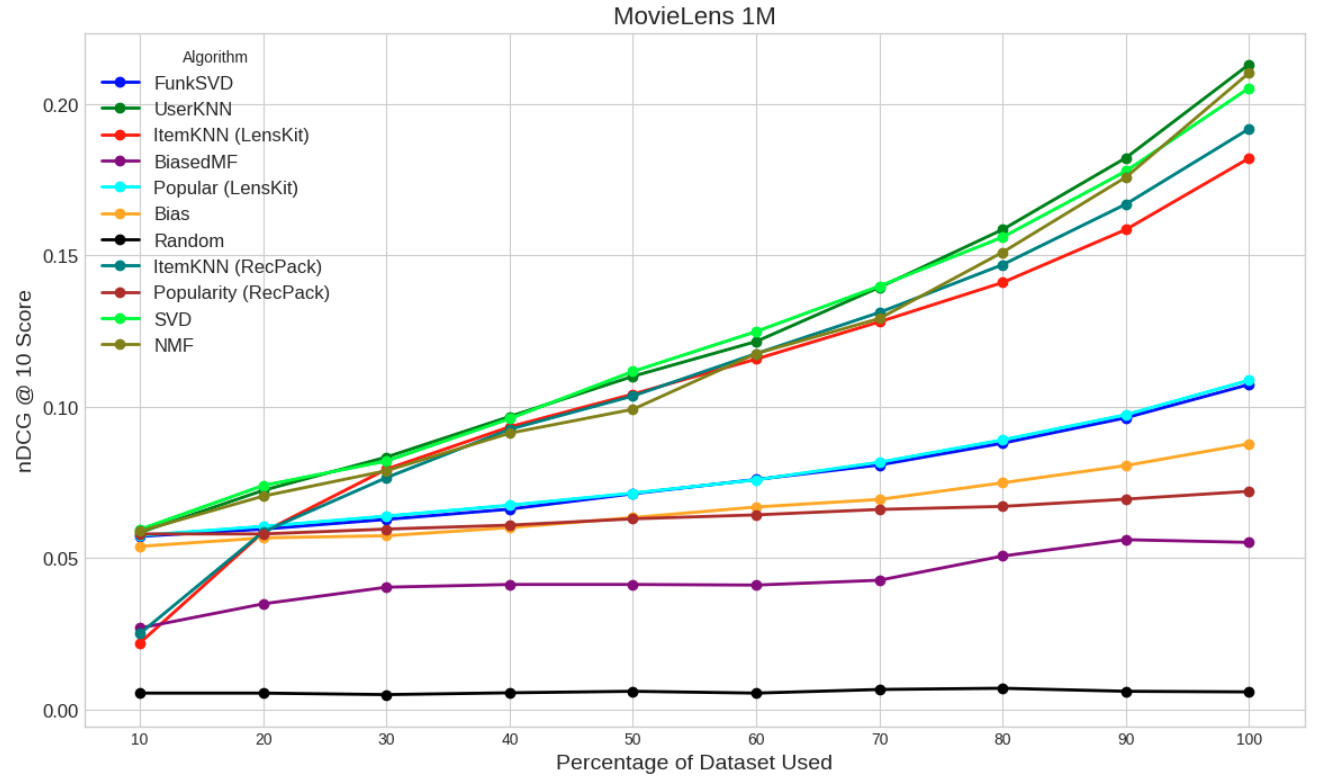}
    \end{minipage}\\
    
    \begin{minipage}[b]{0.48\linewidth}
        \centering
        \includegraphics[width=\linewidth]{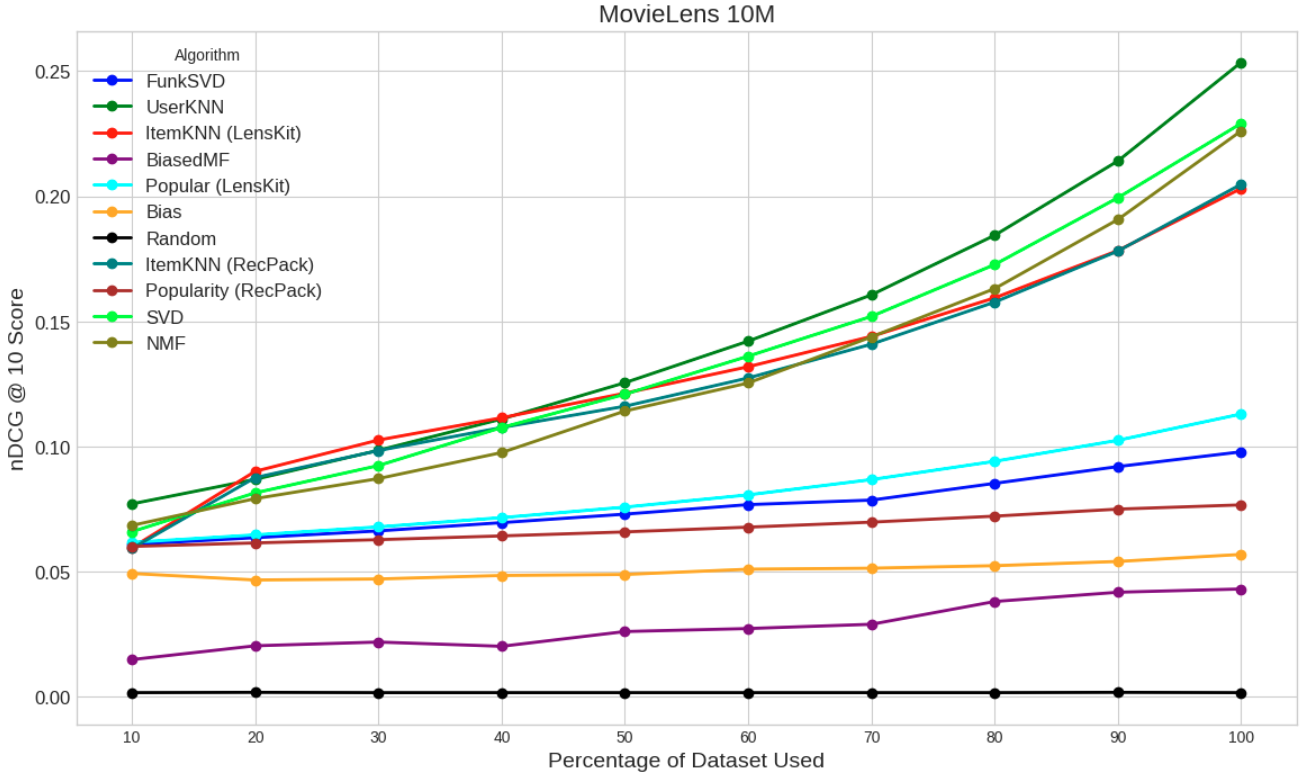}
    \end{minipage}%
    \hfill
    \begin{minipage}[b]{0.48\linewidth}
        \centering
        \includegraphics[width=\linewidth]{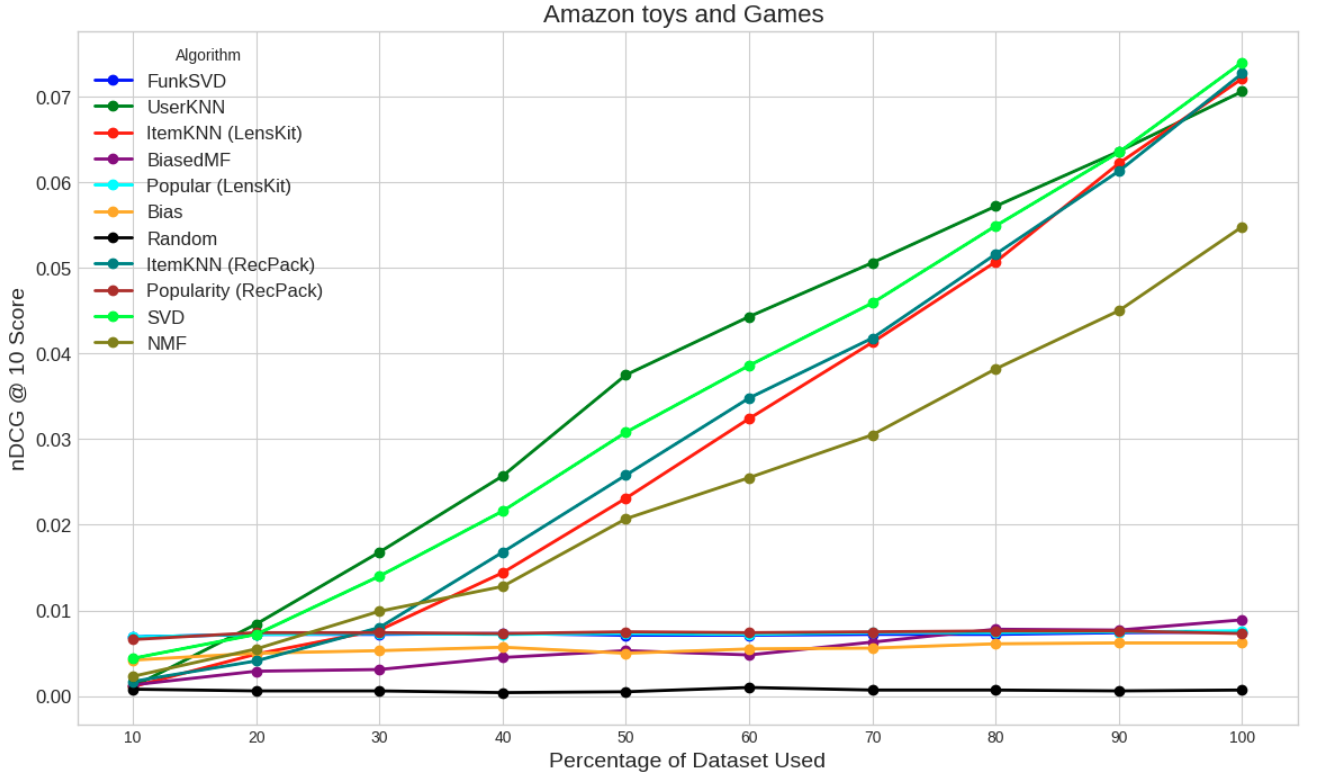}
    \end{minipage}
    \caption{nDCG@10 scores for each algorithm trained on varying portions of the datasets. The horizontal axis shows the percentage of the full training set, where 100\% equals 80\% of the total dataset, with other percentages relative to this.}
    \label{fig:combined-figures}
\end{figure}

Our research investigated the impact of downsampling on the efficiency of recommender system algorithms by analyzing performance metrics and dataset characteristics. Variations in user/item interaction densities and rating distributions, as discussed in subsection 3.1, impact algorithm performance. Preprocessing facilitated consistent evaluations across varying dataset sizes. Before presenting the experimental results (Figure~\ref{fig:combined-figures}), it is useful to estimate the potential environmental benefits of the downsampling strategy proposed in this work, specifically in terms of reducing carbon footprint and CO2e emissions, with a calculation example where the training set is downsampled to 50\% of its original size.

Based on our observations and calculations, downsampling the training data to 50\% reduces the runtime for training and evaluation phases to $\sim$72\% of the runtime required for the full dataset, on average. Furthermore, the energy consumption for a single run of a recommender algorithm on one dataset is estimated at 0.51 kWh\cite{vente2024clicks}. Assuming 10 hyperparameter configurations per algorithm and using the global average conversion factor of 481 gCO2e per kWh\cite{ember2024carbon}, and accounting for a potential increase by a factor of 40 to consider preliminary tasks such as algorithm prototyping, initial tests, debugging, and re-runs\cite{vente2024clicks}, we estimate the potential carbon equivalent emissions savings from downsampling the training set to 50\% compared to the full set per algorithm per dataset as follows:
\[
(100\% - 72\%) \times 0.51 \, \text{kWh} \times 10 \times 481 \, \text{gCO2e/kWh} \times 40 \approx 27.4 \, \text{KgCO2e}.
\]
This estimation roughly quantifies the reduction in CO2e emissions resulting from the training of a single algorithm on a single dataset, based solely on the reduction in runtime following downsampling. It assumes that the hardware used for the full dataset will also be employed for the downsampled dataset and that a nearly linear relationship exists between runtime, energy consumption, and carbon emissions, as supported by the ML CO2 Impact calculator tool\cite{lacoste2019quantifying}.
In the upcoming sections, we detail the principal observations derived from our results and delve into how they inform the objectives of our research. For simplicity in discussing the algorithms examined in this study, we have categorized the algorithms into two groups. This division reflects the observed similarities in performance and results within each group, with distinct behaviors compared to the other group as shown in Figure~\ref{fig:combined-figures}, facilitating clearer analysis of their comparative effectiveness. The Random algorithm serves as a baseline for comparison but is not included in the statistics of either group. Table~\ref{tab:algorithm_groups} provides an overview of these categorizations.\\

\begin{table}[H]
\centering
\caption{Categorization of Examined Algorithms}
\label{tab:algorithm_groups}
\begin{tabular}{|c|p{10cm}|}
\hline
\textbf{Group} & \textbf{Algorithms included} \\
\hline
Group 1 & UserKNN, SVD, ItemKNN (both LensKit and RecPack version), NMF \\ 
\hline
Group 2 & Bias, Popularity, FunkSVD, BiasedMF, Popular \\ 
\hline
\end{tabular}
\end{table}

\subsubsection{\textbf{Observations}} Several key observations can be outlined from our analysis of recommender system algorithms across different datasets, each numbered for easy reference. (1) larger datasets consistently resulted in improved performance across all algorithms, with Group 1 algorithms benefiting significantly from increased data availability. (2) In examining the performance metrics, we observed that Group 1 algorithms displayed significant improvements when the dataset used for training exceeded $\sim$30\% of the total data. Specifically, downsampling the MovieLens 100K dataset to $\sim$50\% resulted in a $\sim$50\% decrease in average nDCG@10 values of this group's algorithms, while reducing to $\sim$30\% led to a $\sim$65\% decrease, highlighting a near-linear relationship between dataset size and performance. (3) Conversely, Group 2 algorithms demonstrated more gradual performance improvements, with nDCG@10 values decreasing by $\sim$23\% and $\sim$29\% in average when the dataset was downsampled to $\sim$50\% and $\sim$30\%, respectively. (4) The sparse Amazon Toys and Games dataset particularly illustrated a more pronounced performance gap between these two groups of algorithms. When downsampling to $\sim$50\% and $\sim$30\%, Group 2 algorithms experienced only $\sim$13\% and $\sim$17\% average drops in performance, respectively, which is less severe compared to the denser MovieLens datasets.\\

\begin{figure}[htbp]
    \centering
    \begin{minipage}[b]{0.24\linewidth}
        \centering
        \includegraphics[width=\linewidth]{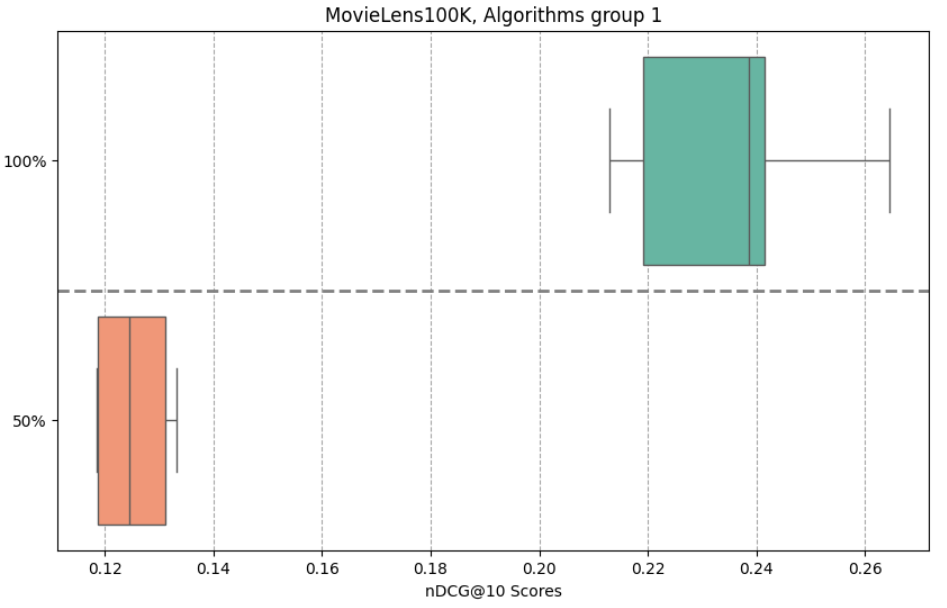}
    \end{minipage}%
    \hfill
    \begin{minipage}[b]{0.24\linewidth}
        \centering
        \includegraphics[width=\linewidth]{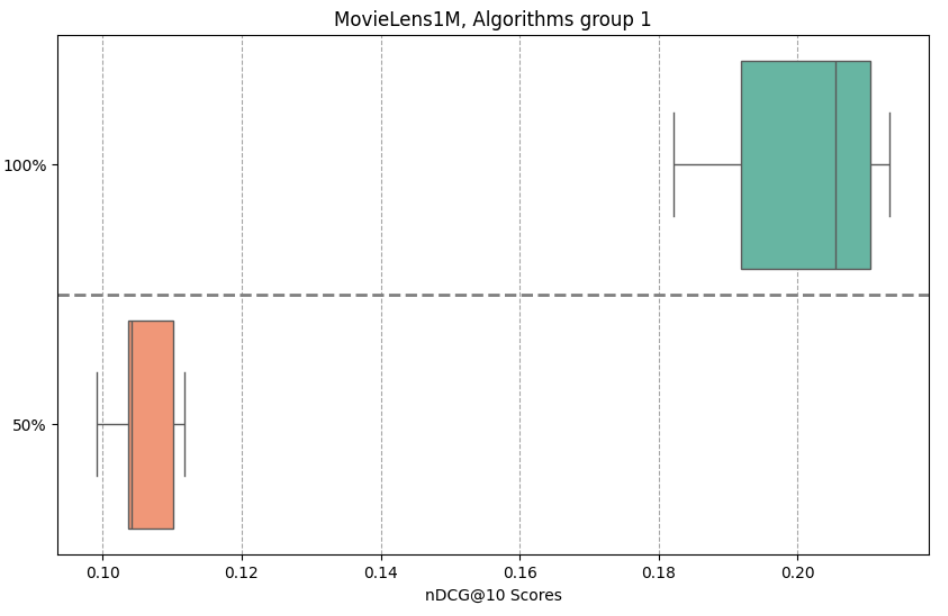}
    \end{minipage}%
    \hfill
    \begin{minipage}[b]{0.24\linewidth}
        \centering
        \includegraphics[width=\linewidth]{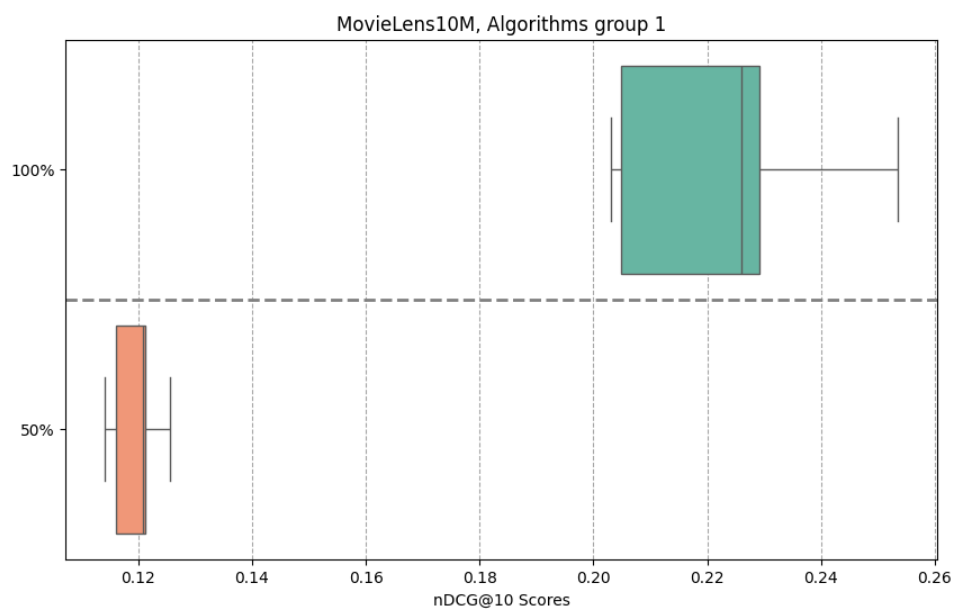}
    \end{minipage}%
    \hfill
    \begin{minipage}[b]{0.24\linewidth}
        \centering
        \includegraphics[width=\linewidth]{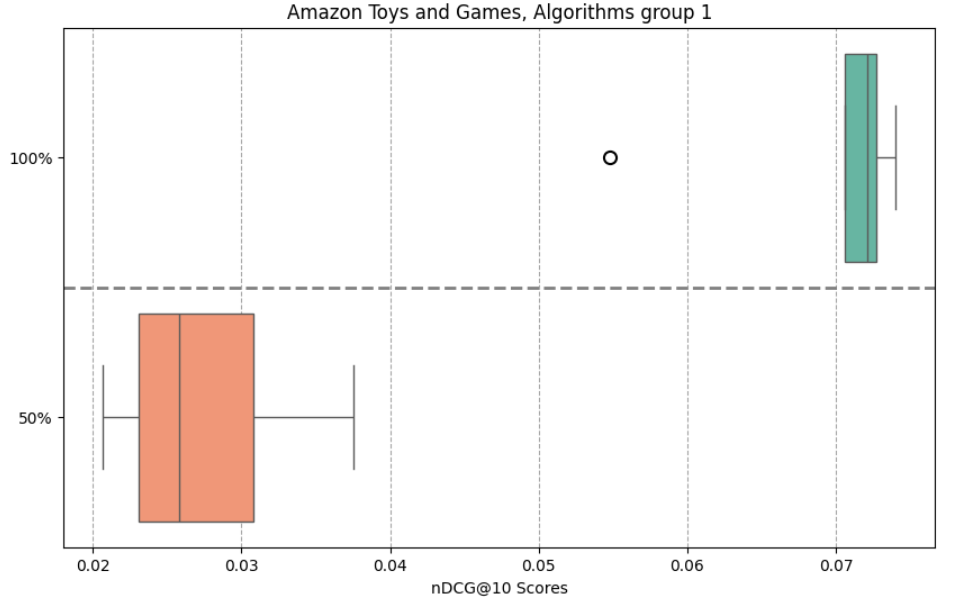}
    \end{minipage}%
    \hfill

    
    \begin{minipage}[b]{0.24\linewidth}
        \centering
        \includegraphics[width=\linewidth]{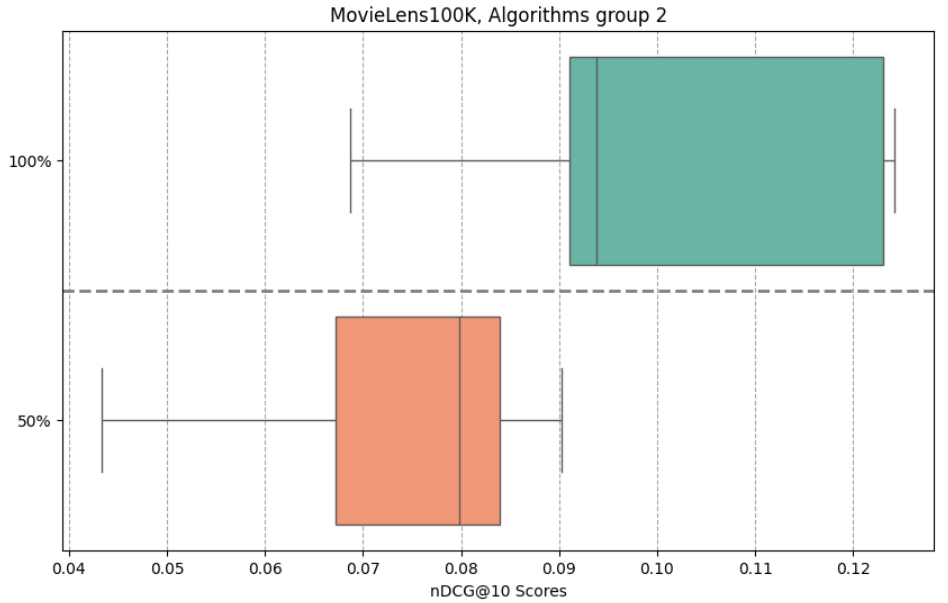}
    \end{minipage}%
    \hfill
    \begin{minipage}[b]{0.24\linewidth}
        \centering
        \includegraphics[width=\linewidth]{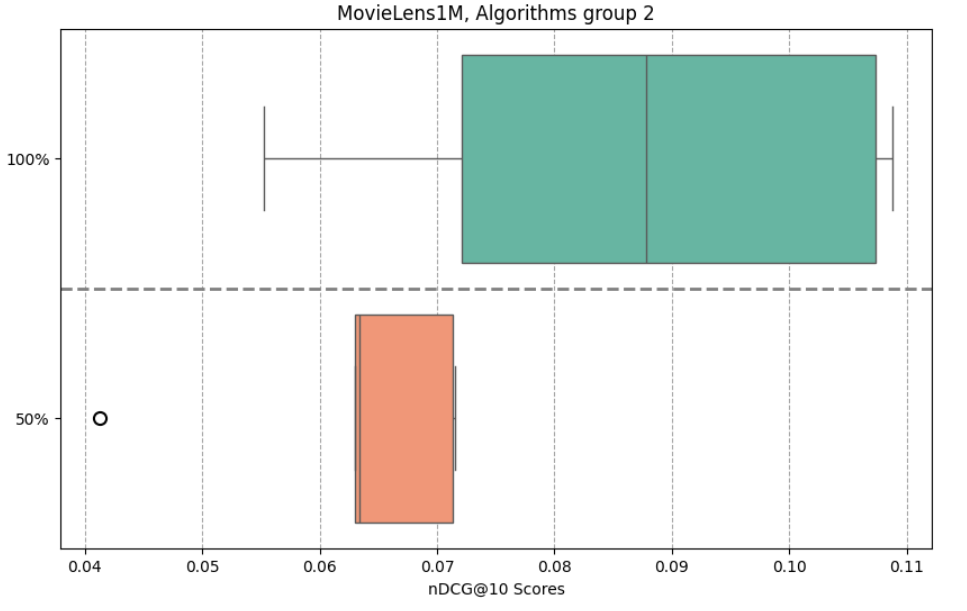}
    \end{minipage}
    \hfill
    \begin{minipage}[b]{0.24\linewidth}
        \centering
        \includegraphics[width=\linewidth]{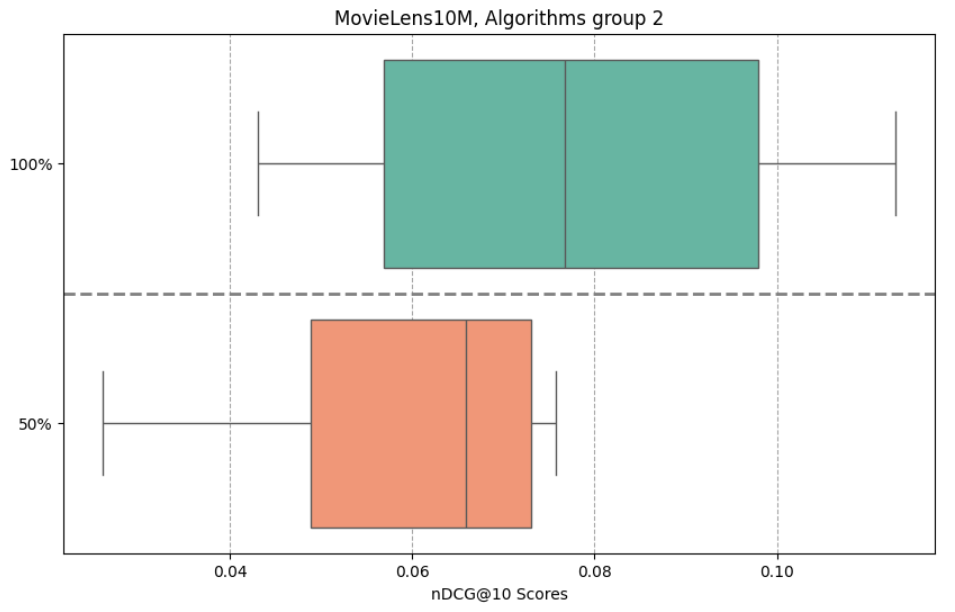}
    \end{minipage}%
    \hfill
    \begin{minipage}[b]{0.24\linewidth}
        \centering
        \includegraphics[width=\linewidth]{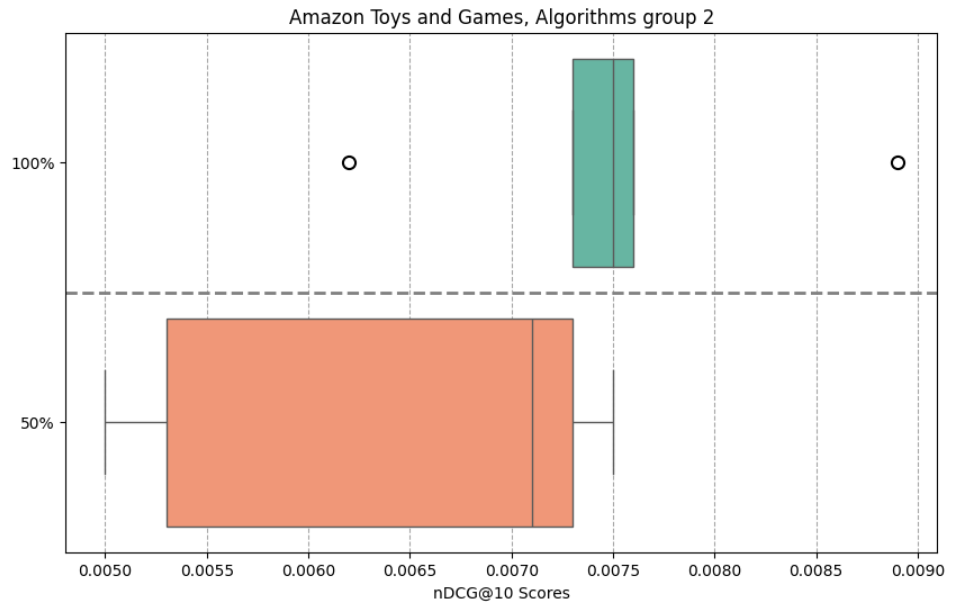}
    \end{minipage}

    \caption{Distribution of nDCG@10 Scores Across Four examined Datasets for Two Distinct Algorithm Groups at 50\% and 100\% Dataset Utilization.}
    \label{fig:overall-label}
\end{figure}

\subsubsection{\textbf{Interpretation}} From these observations, it appears that the size and sparsity of datasets significantly influence the performance of recommender system algorithms. Observation (1) highlights that contrary to our expectations, larger data volumes, including those from extensive datasets like MovieLens 10M, generally lead to better algorithm performance. However, the extent of improvement depends on the specific algorithm and the characteristics of the dataset. Observations (2) and (3) highlight that Group 1 algorithms are highly dependent on larger datasets to perform optimally. In contrast, Group 2 algorithms maintain relatively stable performance even with reduced data, striking a balance between performance and computational efficiency. This observation is evident from the narrower gap in the nDCG@10 scores distribution box plot between 50\% and 100\% dataset utilization for algorithms in Group 2, compared to the larger gap seen in Group 1, as shown in Figure~\ref{fig:overall-label}. The detailed analysis in observation (4) shows that in sparse environments, such as the Amazon Toys and Games dataset, downsampling effectively reduces computational demands with only minimal performance loss. This indicates that strategic downsampling can be a viable method especially in contexts where energy optimization is crucial without significantly compromising accuracy.\\

\subsubsection{\textbf{Conclusion}} This study underscores the potential for optimizing recommender systems through dataset size reduction. Although most algorithms demonstrate enhanced performance with larger training datasets, our analysis has pinpointed specific scenarios where the trade-off between energy efficiency and accuracy favors efficiency. In these cases, significant savings are achieved with minimal detriment to accuracy. Some algorithms consistently maintain high performance even with reduced data volumes, highlighting their potential for energy-efficient AI development.

Therefore, we answer our research question by affirming that it can be possible to identify an optimal trade-off between maintaining algorithmic performance and reducing dataset size. Specifically, our analysis shows that strategic downsampling may improve energy efficiency while maintaining performance comparable to the original dataset size, thereby supporting the optimization of AI systems and recommenders. However, more research is necessary to find out when exactly downsampling is a sensible approach, as sometimes, performance varies notably. We hope that in the long term, downsampling datasets becomes an accepted best-practice \cite{Beel2024,Beel2024d}, for the recommender-system community that helps to contribute to green and sustainable recommender systems. \\

\subsubsection{\textbf{Acknowledgment}} This paper benefited from ChatGPT for grammar and wording improvements\cite{beel2024ai}. The code used for conducting the experiments is publicly available on GitHub\cite{recso2024github}.

\bibliographystyle{splncs04}
\bibliography{references}

\end{document}